\documentclass{article}

\usepackage[table]{xcolor}
\usepackage{amssymb}
\usepackage{amsfonts}
\usepackage{amsmath}
\usepackage{booktabs}
\usepackage{caption}
\usepackage{microtype}
\usepackage{nicefrac}
\usepackage{graphicx}
\usepackage{subfigure}
\usepackage{tikz}
\usetikzlibrary{decorations.text}
\usepackage{enumitem}
\usepackage{pgfplots}
\usepackage{caption} 
\usepackage{xfrac}

\definecolor{mygray}{gray}{0.5}

\usepackage{hyperref}

\usepackage[accepted]{icml2020}

\usetikzlibrary{arrows,shapes.geometric,positioning}
\pgfplotsset{compat=newest}

\DeclareMathOperator*{\argmax}{arg\,max}
\newcommand \mask{\varnothing}

\newcommand\vertarrowbox[3][0.0ex]{%
  \begin{array}[t]{@{}c@{}} #2 \\
  \left\uparrow\vcenter{\hrule height #1}\right.\kern-\nulldelimiterspace\\
  \makebox[0pt]{\scriptsize#3}
  \end{array}%
}

\icmltitlerunning{Imputer: Sequence Modelling via Imputation and Dynamic Programming}

\begin{document}

\twocolumn[
\icmltitle{Imputer: Sequence Modelling via Imputation and Dynamic Programming}



\icmlsetsymbol{resident}{\dag}
\icmlsetsymbol{doneatgoogle}{\ddag}

\begin{icmlauthorlist}
\icmlauthor{William Chan}{brain}
\icmlauthor{Chitwan Saharia}{brain,resident}
\icmlauthor{Geoffrey Hinton}{brain}
\icmlauthor{Mohammad Norouzi}{brain}
\icmlauthor{Navdeep Jaitly}{deshaw}
\end{icmlauthorlist}

\icmlaffiliation{brain}{Google Research, Brain Team, Toronto, Ontario, Canada.}
\icmlaffiliation{deshaw}{Work done at Google; currently at The D. E. Shaw Group, New York, New York, USA}

\icmlcorrespondingauthor{W. Chan}{\mbox{williamchan@google.com}}
\icmlcorrespondingauthor{C. Saharia}{sahariac@google.com}
\icmlcorrespondingauthor{G. Hinton}{geoffhinton@google.com}
\icmlcorrespondingauthor{M. Norouzi}{mnorouzi@google.com}
\icmlcorrespondingauthor{N. Jaitly}{navdeep.jaitly@deshaw.com}

\icmlkeywords{Machine Learning, Deep Learning, Speech Recognition}

\vskip 0.3in
]



\printAffiliationsAndNotice{\textsuperscript{\dag}Work done as part of the Google AI Residency.}  

\begin{abstract}
This paper presents the Imputer, a neural sequence model that generates output sequences iteratively via imputations. The Imputer is an iterative generative model, requiring only a constant number of generation steps independent of the number of input or output tokens. The Imputer can be trained to approximately marginalize over all possible alignments between the input and output sequences, and all possible generation orders. We present a tractable dynamic programming training algorithm, which yields a lower bound on the log marginal likelihood. When applied to end-to-end speech recognition, the Imputer outperforms prior non-autoregressive models and achieves competitive results to autoregressive models. On LibriSpeech test-other, the Imputer achieves 11.1 WER, outperforming CTC at 13.0 WER and seq2seq at 12.5 WER.
\end{abstract}

\vspace*{-.3cm}
\section{Introduction}
\vspace*{-.1cm}

Neural sequence models have been widely successful across a wide range of applications, including machine translation \citep{bahdanau-iclr-2015,luong-emnlp-2015}, speech recognition \citep{graves-icml-2014,chan-icassp-2016}, speech synthesis \citep{oord-arxiv-2016,oord-icml-2018} and image captioning \citep{vinyals-cvpr-2015,xu-icml-2015}. Autoregressive sequence models (e.g., \citet{sutskever-nips-2014,cho-emnlp-2014}) enable exact likelihood estimation, at the cost of requiring $n$ generation steps to generate $n$ tokens during inference. On the other hand, non-autoregressive models such as CTC \citep{graves-icml-2006} and NAT \citep{gu-iclr-2018} can generate sequences in a single generation step, independent of the number of output tokens. However these non-autoregressive models typically make a strong conditional independence assumption between output tokens, often underperforming their autoregressive counterparts. Recently, there has been a growing interest in models that make a trade-off between the two extremes of fully autoregressive and fully non-autoregressive generation, such as the Insertion Transformer \citep{stern-icml-2019}, Mask-Predict \citep{ghazvininejad-emnlp-2019}, Levenstein Transformer \citep{gu-neurips-2019} and Multilingual KERMIT \cite{chan-neurips-2019}. Such models sacrifice almost no performance, while requiring a logarithmic \citep{chan-arxiv-2019b} or a constant number of generation steps \cite{lee-emnlp-2018}.

In this paper, we seek to extend prior work to achieve a desirable balance between
fully autoregressive and fully non-autoregressive models. We are concerned with sequence problems in which a natural monotonic latent alignment exists between the source and target sequences (e.g., end-to-end speech recognition). Prior work typically does not take advantage of such useful inductive biases, which become increasingly important for problems with long output sequences. For instance, in speech recognition sequences are typically an order of magnitude longer than sequences seen in machine translation. The need to generate longer sequences highlights the importance of developing sequence generation frameworks with sub-linear inference time.

Three popular types of generative models have been developed for end-to-end speech recognition applications: CTC \citep{graves-icml-2006,graves-icml-2014}, RNN-T \citep{graves-icml-2012,graves-icassp-2013} and seq2seq \citep{chan-icassp-2016,bahdanau-icassp-2016}. CTC is a non-autoregressive model and makes a conditional independence assumption between token frame predictions. Accordingly, CTC can make use of dynamic programming to marginalize over all possible alignments. CTC models exhibit fast inference but relatively poor performance \cite{battenberg-asru-2017}.
RNN-T is an autoregressive model that captures conditional dependencies between output tokens, but at the cost of $n$ generation steps to generate $n$ tokens. Similar to CTC, the RNN-T relies on dynamic programming to marginalize over all possible alignments for exact likelihood computation. However, the RNN-T dynamic programming framework typically relies on a linear bottleneck for tractability (e.g., the linear bottleneck between the prediction and transcription network \citep{graves-icml-2012}); this has typically limited the empirical success of RNN-T \citep{prabhavalkar-interspeech-2017}.
Seq2seq is an autoregressive left-to-right model, which factorizes the probability with the chain rule resulting in exact likelihood estimates.
Empirically, seq2seq often performs the best \citep{chiu-icassp-2018}. However, the seq2seq framework requires $n$ generation steps to generate $n$ tokens.

\begin{figure}[t]
\centering
    \resizebox{\columnwidth}{!}{
\begin{tikzpicture}
\tikzstyle{imputer} = [rectangle, thick, rounded corners, minimum width=2.0cm, minimum height=1cm,text centered, draw=black]
\tikzstyle{symbol} = [align=center, minimum height=0.6cm]
\tikzstyle{arrow} = [thick,->]

\node[fill=white,opacity=.2,minimum width=8.8cm,minimum height=1.3cm,inner sep=0cm] (background0) at (0,1.8) {};
\node[minimum width=6cm] (step0) at (0, 1.55) {
\begin{tabular}{|p{0.2cm}|p{0.2cm}|p{0.2cm}!{\vrule width 1.5pt}p{0.2cm}|p{0.2cm}|p{0.2cm}!{\vrule width 1.5pt}p{0.2cm}|p{0.2cm}|p{0.2cm}!{\vrule width 1.5pt}p{0.2cm}|p{0.2cm}|p{0.2cm}|}
 \hline
 $\varnothing$ & $\varnothing$ & $\varnothing$ & $\varnothing$ & $\varnothing$ & $\varnothing$ & $\varnothing$ & $\varnothing$ & $\varnothing$ & $\varnothing$ & $\varnothing$ & $\varnothing$ \\
 \hline
 \end{tabular}
 };
\node[text width=8cm,align=justify] (description0) at (0,2.1) {1. Initial alignment is filled with mask tokens $\varnothing$.};

\node (imputer) [imputer] at (5.2, -1.6) {Imputer};

\node[fill=white,opacity=.2,minimum width=8.8cm,minimum height=1.3cm,inner sep=0cm] (background1) at (0, 0) {};
\node[minimum width=6cm] (step1) at (0, -0.25) {
\begin{tabular}{|p{0.2cm}|p{0.2cm}|p{0.2cm}!{\vrule width 1.5pt}p{0.2cm}|p{0.2cm}|p{0.2cm}!{\vrule width 1.5pt}p{0.2cm}|p{0.2cm}|p{0.2cm}!{\vrule width 1.5pt}p{0.2cm}|p{0.2cm}|p{0.2cm}|}
 \hline
 $A$ & $\varnothing$ & $\varnothing$ & $\varnothing$ & $\_$ & $\varnothing$ & $\varnothing$ & $E$ & $\varnothing$ & $\varnothing$ & $\varnothing$ & $F$ \\
 \hline
 \end{tabular}
};
\node[text width=8cm,align=justify] (description1) at (0,0.3) {2. Imputer conditions on a previous alignment.};

\node[fill=white,opacity=.2,minimum width=8.8cm,minimum height=2cm,inner sep=0cm] (background2) at (-0.0cm, -3.5cm) {};
\node[minimum width=6cm] (step2) at (0, -4.0) {
\begin{tabular}{|p{0.2cm}|p{0.2cm}|p{0.2cm}!{\vrule width 1.5pt}p{0.2cm}|p{0.2cm}|p{0.2cm}!{\vrule width 1.5pt}p{0.2cm}|p{0.2cm}|p{0.2cm}!{\vrule width 1.5pt}p{0.2cm}|p{0.2cm}|p{0.2cm}|}
 \hline
  $A$ & $B$ & $\varnothing$ &    $\varnothing$ & $\_$ & $C$ &    $D$ & $E$ & $\varnothing$ &     $\varnothing$ & $\_$ & $F$\\
 \hline
 \end{tabular}
};
\node[text width=8cm,align=justify] (description2) at (0.0,-2.8) {3. Imputer generates a new alignment. For each block, the token with the largest probability is selected and merged with the existing alignment.};

\draw[arrow] (background2.west) to [bend left=90] node[above ,sloped] {Repeat $B$ iterations} (background1.west);
\draw[arrow] (background1.east)  to [out=0,in=90,looseness=1]  (imputer.north) ;
\draw[arrow] (imputer.south)  to [out=270,in=0,looseness=1]  (background2.east) ;
\draw[arrow] (background0.south) to (background1.north) ;

\end{tikzpicture}}
\caption{Visualization of the Imputer's decoding procedure. For this example, the alignment comprises $4$ blocks with block size of $B=3$ tokens each. The alignment ``A B \_ \_ \_ C D E \_ \_ \_ F'' is being imputed, with 1 token per block imputed at each decoding iteration. The decoding process takes exactly $B$ iterations.}
\label{fig:imputer_decoding}
\end{figure}
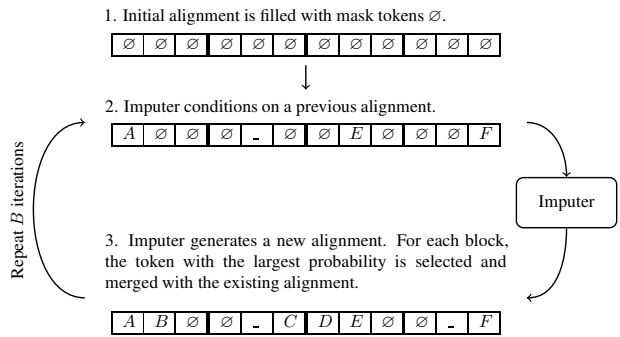

This paper presents the Imputer, with the following important characteristics:
\begin{enumerate}[nosep]
    \item Imputer is an iterative generative model, requiring a constant number of generation steps independent of the number of output sequence tokens (see Figure~\ref{fig:imputer_decoding}).
    \item Imputer approximately marginalizes over all possible alignments and generation orders. It is well suited for problems with a latent monotonic alignment such as speech recognition.
    \item Imputer is not left-to-right. It can model language with bidirectional contextualization, and can model both local and global conditional dependencies.
    \item Imputer superimposes the input and output sequences together, which allows the Imputer to simplify the architecture and avoid the cross-attention mechanism typically found in encoder-decoder sequence models.
\end{enumerate}
\section{The Imputation Framework}
\label{sec:imputation_framework}

{\textbf{Notation.}}
Let $x$ denote the input sequence (e.g., audio) and $y$ denote the output sequence (e.g., text). Let $y_i \in \mathcal{V}$ denote a token in the output sequence, where $\mathcal{V}$ is the vocabulary (e.g., alphabet).
It is not required that $x$ and $y$ have an equal length, but $x$ must be longer than or equal in length to $y$ (i.e., $|x| \geq |y|$).
Let $a$ denote a latent alignment between $x$ and $y$, with $a_i \in \mathcal{V}^+$ where $\mathcal{V}^+ = \mathcal{V} \cup \{``\_"\}$ and ``$\_$'' is a special token called the blank token, which helps align $x$ and $y$.
This blank token is compatible with the definition from CTC~\citep{graves-icml-2006}.
We assume there exists a monotonic alignment between $x$ and $y$, and consequently $|a| = |x| \geq |y|$, such that when we remove the blank tokens from $a$, we recover $y$. 
We also introduce the function $\beta(y)$ that returns the set of all possible alignments for $y$ (with length $|x|$). Let $\beta^{-1}(a)$ denote the collapse function which returns the $y$ associated with a given $a$.
Here is a concrete example: let $y = (A,B,C,D)$ and $|x| = 7$, then, one possible target alignment is $a=(\_,A,B,\_,C,\_,D)$ and $y = \beta^{-1}(a)$.
We note to an astute reader well versed with CTC that the discussion of collapsing repetitions typically found in the CTC literature \cite{graves-icml-2006,graves-icml-2014} is omitted, but CTC and Imputer can handle both cases, and this is simply a hyperparameter choice.

\subsection{Connectionist Temporal Classification (CTC)}
We will first briefly discuss Connectionist Temporal Classification (CTC) \citep{graves-icml-2006}. CTC models an alignment $a$ with the conditional independence assumption between token frame predictions:
\begin{align}
    p_\theta(a | x) = \prod_i p(a_i | x; \theta)
\end{align}
CTC is typically trained to maximize the log-likelihood:
\begin{align}
    \log p_\theta(y|x) = \log \sum_{a \in \beta(y)} p_\theta(a | x) 
    \label{eqn:ctc_dynamic_programming}
\end{align}
Equation \ref{eqn:ctc_dynamic_programming} marginalizes over all possible alignments $\beta(y)$ compatible with $y$. The conditional independence assumption warrants Equation \ref{eqn:ctc_dynamic_programming} to be solved exactly and efficiently via dynamic programming \citep{graves-icml-2006}, and permits for fully non-autoregressive generation in one inference step. However, the conditional independence assumption also means tokens are generated without dependencies on each other, consequently the model cannot easily model language dependencies and/or multimodal outputs.

\subsection{The Imputer}

Imputer is an iterative generative model. At each generative step, Imputer conditions on a previous partially generated alignment and emits a new alignment. Each generative step emits a complete alignment, but, during inference only a small subset is selected to generate the next partial alignment. This subset typically includes the previously emitted tokens, resulting in an iterative refinement process.
This iterative process allows Imputer to model conditional dependencies across generation steps, which CTC lacks. The Imputer parameterizes the alignment distribution with a conditional independence assumption between token predictions:
\begin{align}
    p_\theta(a | \tilde a, x) = \prod_i p(a_i | \tilde a, x; \theta) \label{eqn:factorization}
\end{align}
where $\tilde a$ is some previous alignment, and $\tilde a_i \in \mathcal{V}^{++}$ where $\mathcal{V}^{++} = \mathcal{V}^+ \cup \{\mask\}$ where $\mask$ is the masked out token. The $\mask$ mask token is compatible with the mask token definition from BERT \citep{devlin-naacl-2019}. To be explicit, $\tilde a$ may be all $\mask$ (e.g., all tokens are masked out), and $a_i = \tilde a_i \forall \tilde a_i \neq \mask$ (i.e., once a token is committed in $\tilde a$, $a$ must be consistent). Here is a concrete example, if $y = (A,B,C,D)$ and $|x| = 7$, given a previous alignment $\tilde a=(\mask,A,\mask,\mask,C,\_,D)$, one possible target alignment is $a=(\_,A,B,\_,C,\_,D)$.
 One key property of Imputer is that, it can model conditional dependencies through the conditioning on $\tilde a$ (e.g., across generation steps), and achieve parallel generation through the conditional independence assumption in-between new token predictions (e.g., within a generation step).

\subsection{Inference}
\label{sec:inference}

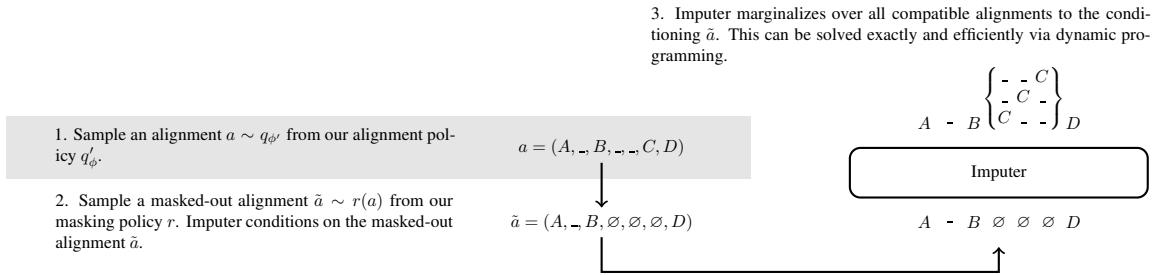
\begin{figure*}[t]
\centering
\begin{tikzpicture}[node distance=1cm, scale=0.66, every node/.style={transform shape}]
\tikzstyle{imputer} = [rectangle, thick, rounded corners, minimum width=6cm, minimum height=1cm,text centered, draw=black]
\tikzstyle{symbol} = [align=center, minimum height=0.6cm]
\tikzstyle{arrow} = [thick,->]

\node[fill=gray,opacity=.2,minimum width=15cm,minimum height=1.25cm,inner sep=0cm] (background1) at (-12.5,0) {};
\node[minimum width=6cm] (step1) at (-8, 0) {$a = (A,\_,B,\_,\_,C,D)$};
\node[text width=8cm,align=justify] (description1) at (-15,0) {1. Sample an alignment $a \sim q_{\phi'}$ from our alignment policy $q_\phi'$.};

\node[minimum width=6cm] (step2) at (-8,-1.5) {$\tilde a = (A,\_,B,\varnothing,\varnothing,\varnothing,D)$};
\node[text width=8cm,align=justify] (description2) at (-15,-1.5) {2. Sample a masked-out alignment $\tilde a \sim r(a)$ from our masking policy $r$. Imputer conditions on the masked-out alignment $\tilde a$.};

\draw[arrow] (step1.south) -- (step2.north);
\draw[arrow] (step2.south) -- (-8,-2.5) -- (0,-2.5) -- (0, -2);

\node[text width=10cm,align=justify] (description3) at (-2,2.25) {3. Imputer marginalizes over all compatible alignments to the conditioning $\tilde a$. This can be solved exactly and efficiently via dynamic programming.};

\node (imputer) [imputer] at (0,-0.5) {Imputer};

\node[symbol] at (-1.5, -1.5) {$A$};
\node[symbol] at (-1.0, -1.5) {$\_$};
\node[symbol] at (-0.5, -1.5) {$B$};
\node[symbol] at (-0.0, -1.5) {$\varnothing$};
\node[symbol] at (0.5, -1.5) {$\varnothing$};
\node[symbol] at (1.0, -1.5) {$\varnothing$};
\node[symbol] at (1.5, -1.5) {$D$};

\node[symbol] at (-1.5, 0.5) {$A$};
\node[symbol] at (-1.0, 0.5) {$\_$};
\node[symbol] at (-0.5, 0.5) {$B$};
\node[symbol] at (0.5, 1.0) {$\left \{ \begin{array}{@{}c@{\hspace*{.1cm}}c@{\hspace*{.1cm}}c@{}}
\_ & \_ & C \\
\_ & C & \_ \\
C & \_ & \_
\end{array}
\right \}$};
\node[symbol] at (1.5, 0.5) {$D$};

\end{tikzpicture}
\caption{Visualization of the Imputer dynamic programming training procedure. A roll-in policy is used to sample a masked-out alignment $\tilde a$. Imputer marginalizes over all compatible alignments with $\tilde a$ over the masked-out regions.}
\label{fig:dynamic_programming}
\end{figure*}

We will first describe inference of Imputer, then we will describe the training objective function afterwards.
The structure of the Imputer framework allows for very flexible balance between autoregressive and non-autoregressive generation. On one extreme, Imputer can be fully non-autoregressive with parallel generation, here $\tilde a$ is simply all $\mask$ blank tokens (i.e., not conditioned on any information), and $a$ and consequently $y$ is generated in just 1 step. This would make full conditional independence assumption between tokens, falling back to CTC-like model \citep{graves-icml-2006} (depending on the loss chosen to train Imputer).

On the other extreme, Imputer can be fully autoregressive with serial generation: At each generation step, we sample one token from Imputer (e.g., Equation \ref{eqn:factorization}), join it with the previous hypothesis $\tilde a$, and repeat this process until $\tilde a$ has no more masked tokens. This autoregressive inference process will take exactly $|x|$ generation steps. This procedure will not make any conditional independence assumption between tokens, but at the cost of generation speed. Note, this autoregressive process need not be left-to-right, as there is no constraint on the generation order.

We present an alternative inference procedure that takes a balance between fully autoregressive generation and fully non-autoregressive generation. Our inference procedure will result in a constant number of generation steps, while still modelling conditional dependencies on both a local and global basis.
Our inference strategy begins by segmenting the sequence $a$ into equal sized blocks of size $B$. We initialize a previous alignment with all masked out tokens $\mask$. We then compute the model's new alignment prediction (conditioned on the source sequence and the previous alignment). For each block, we select the slot (and its token) with the maximum token-level probability prediction from the model (among slots that are masked out). We merge this prediction with the previous alignment. This partially predicted alignment is then used for inference in the next decoding iteration. Since we use fixed and equal sized blocks, and each decoding iteration ensures exactly one token imputation per block, the inference procedure will finish in exactly $B$ iterations. Note that in the extreme case of $B = 1$, this will result in fully non-autoregressive generation, while the other extreme $B =|a|$ will result in fully autoregressive generation.  

\subsection{Marginalization}

We now proceed to describe the objective function of Imputer. We start off by writing the marginalization over all possible alignments and generation orders:
\begin{align}
    & \log p_\theta(y | x) = \log \sum_{a \in \beta(y)} \sum_{\tilde a \in \gamma(a)} p_\theta(a | \tilde{a}, x) p(\tilde a | x) \label{eqn:summation}
\intertext{
where $\beta(y)$ captures all possible alignments of $y$, $\gamma(a)$ captures all possible masking permutations, and $p(\tilde a | x)$ is some prior. The combination of $\beta \times \gamma$ will capture all possible alignments and masking permutations, and consequently it will capture all possible alignments and generation order. We can rewrite this under expectation of the importance sampling distributions $q$ and $r$:
}
    &\quad= \log \mathbb{E}_{a \sim q} \left[ \frac{1}{q(a)} \mathbb{E}_{\tilde a \sim r} \left[ \frac{1}{r(\tilde a)} p_\theta(a | \tilde{a}, x) p(\tilde a | x) \right] \right] \label{eqn:importance_sampling} \\
    &\quad\geq \mathbb{E}_{a \sim q} \left[ \mathbb{E}_{\tilde a \sim r} \left[ \log p_\theta(a | \tilde{a}, x) + \log p(\tilde a | x) \right] + H(r) \right] + H(q) \\
    &\quad\geq \mathbb{E}_{a \sim q} \left[ \mathbb{E}_{\tilde a \sim r} \left[ \log p_\theta(a | \tilde{a}, x) \right] + H(r) \right] + H(q) \label{eqn:marginalization}
\end{align}
where $q$ replaces $\beta$ to capture all possible alignments, and $r$ replaces $\gamma$ to capture all possible maskings. As long as there is support everywhere under $q$ and $r$, this equality in Equation \ref{eqn:importance_sampling} holds. We can further use Jensen's inequality to create a lower bound in Equation \ref{eqn:marginalization}. Assuming $q$ and $r$ are distributions that do not share parameters with $p_\theta$, we can ignore the entropy terms during the gradient computation. We define our objective function based on Equation \ref{eqn:marginalization} without the constant entropy terms:
\begin{align}
    J(\theta) = \underbrace{\mathbb{E}_{a \sim q} [ \mathbb{E}_{\tilde a \sim r}}_\text{roll-in policy} [ \log p_\theta(a | \tilde{a}, x) ] ] \label{eqn:objective}
\end{align}
There are two main problems with Equation \ref{eqn:objective}, 1) high variance due to the double sampling procedure, and 2) it is unclear how to choose $r$ and perhaps more importantly how to choose $q$. Consequently, naively optimizing for Equation \ref{eqn:objective} directly is difficult in practice \citep{luo-icassp-2016}.
Equation \ref{eqn:objective} can be interpreted as $q \times r$ being our roll-in policy to generate a $\tilde a$, and our model $\theta$ is taught to complete the alignment output given some sampled previous alignment $\tilde a$. One way to reinterpret the training problem is, 1) what should the roll-in policy used to sample $\tilde a$ be? and 2) given a partially completed alignment $\tilde a$, what should the target policy be to complete the alignment? We will discuss the choice of roll-in policy in more detail in Section \ref{sec:rollin_policy}. We take two approaches for our target policy, 1) we use Imitation Learning and follow an expert, and 2) we use Dynamic Programming to marginalize over all possible alignments compatible with $\tilde a$.  We will discuss these target policies in detail in Section \ref{sec:imitation_learning} and \ref{sec:dynamic_programming}.

\subsection{Roll-in Policy}
\label{sec:rollin_policy}

The ideal choice of the roll-in policy is obvious, we want to align our training and inference conditions, thus the ideal choice of the roll-in should be drawing $\tilde a$ samples from our model's policy $p_\theta$. However, this would be prohibitively expensive as it would involve sampling/decoding on-the-fly during training. Our strategy is to sample from some other distribution, which is inexpensive yet close to our model $\theta$.

\textbf{Alignment Policy $\beta$}. We will first discuss how to sample alignments. Let us assume we had access to an expert $\phi$ (e.g., existing CTC or Imputer model). We could sample $q_\phi$ and use this as our alignment sampling distribution. We could in theory compute the logits for all the examples in the training distribution, store them offline, and sample from it during training. However, this would be prohibitively expensive as it would require large amounts of memory. We found it convenient and working well in practice to simply create a pseudo-expert policy $q_{\phi'}$ as a distribution we can easily sample from. We first search for the best empirical alignment $\hat a_\phi^*$ under the expert $\phi$:
\begin{align}
    \hat a_\phi^* &= \underset{a}{\argmax} \, q_\phi(a | x, y) \label{eqn:argmax_alignment}
\end{align}
For certain classes of models (e.g., CTC), Equation \ref{eqn:argmax_alignment} can be solved efficiently and exactly via dynamic programming \cite{graves-icml-2006}. We then create $q_{\phi'}$ from $\hat a_\phi^*$ by adding some small noise:
\begin{align}
    q_{\phi'} = \hat a_\phi^* + \mathcal{N} \label{eqn:expert_noise}
\end{align}
where $\mathcal N$ is some noise distribution (e.g., randomly shifting the alignment left-or-right). Sampling from $q_{\phi'}$ is especially convenient because we can store $\hat a_\phi^*$ offline, and sample the noise distribution $\mathcal{N}$ on-the-fly during training (which is cheap), and there would be no need to store or sample from $q_\phi$ directly during training.

An alternative approach is to draw samples from $q_{\theta'}$, a stationary distribution created from a stale copy of the model $\theta'$. This approach would more closely align the training and inference conditions. This approach is also convenient since we will not need to do live decoding or sampling during training, and since $q_{\theta'}$ is stationary, we also would not need to backpropagate through it. We experiment with both approaches in our experimental section.

\textbf{Masking Policy $\gamma$.} Until now, we have only discussed how to select the $q$ part of the roll-in alignment policy, we will now discuss our masking policy $r$. There are obvious choices for $r$, for example Bernoulli \citep{devlin-naacl-2019} or Uniform distributions \citep{stern-icml-2019}. We also designed a masking policy inline with our inference procedure described in Section \ref{sec:inference}. During inference, we $\argmax$ one token from each block, parallel across all blocks. This means at any given point, there is an equal number of tokens per block. Following this observation, we design our Block sampling strategy to follow this inference condition. We first sample a number $b \in [0, B)$, where $B$ is the size of our block. We then simply randomly mask out $b$ number of tokens per block.

\subsection{Imitation Learning}
\label{sec:imitation_learning}

Let us assume we had access to the same (psuedo-)expert $\phi'$ (as described in Section \ref{sec:rollin_policy}). One simple objective is to simply imitate the expert:
\begin{align}
    J_{\textrm{IM}}(\theta) = {\mathbb{E}_{a \sim q_{\phi'}}} \left[ \mathbb{E}_{\tilde a \sim r} \left[ \log p_\theta(a | \tilde{a}, x) \right] \right] \label{eqn:efficient_imitation_learning}
\end{align}
Equation \ref{eqn:efficient_imitation_learning} can be interpreted as $q_{\phi'} \times r$ being our roll-in policy, and our model $\theta$ is taught to imitate the pseudo-expert $\phi'$.

\subsection{Dynamic Programming}
\label{sec:dynamic_programming}

One limitation of the Imitation Learning approach above is that when we distill $q_{\phi'}$ as our target distribution, we are constrained to the knowledge (and errors) of $\phi'$. This is especially concerning if $\phi'$ does not capture the best possible alignment for our model $\theta$. We present a new objective function which decouples the role of $q_{\phi'}$ as both the roll-in distribution and target policy. We keep $q_{\phi'} \times r$ as our roll-in distribution, but for our target policy we marginalize over all possible compatible alignments:
\begin{align}
J_{\textrm{DP}}(\theta) &= \mathbb{E}_{a \sim q_{\phi'}} \left[ \mathbb{E}_{\tilde a \sim r} \left[ \log \sum_{a' \in \beta'(\tilde a, a)} p_\theta(a' | \tilde{a}, x) \right] \right] \label{eqn:dynamic_programming}
\end{align}
where $\beta'(\tilde a, a)$ returns the set of all possible alignments compatible with $\tilde a$ drawn from our roll-in distribution $q_{\phi'} \times r$. The key difference between Equation $\ref{eqn:efficient_imitation_learning}$ and Equation $\ref{eqn:dynamic_programming}$ is the extra summation $\sum_{a' \in \beta'(\tilde a, a)}$. The summation says, under the roll-in distribution of $q_{\phi'} \times r$, let us marginalize over all compatible alignments with $\tilde a$. Conveniently, the $\log \sum$ term in Equation \ref{eqn:dynamic_programming} can be computed exactly and efficiently via dynamic programming \citep{graves-icml-2006}. The dynamic programming to solve Equation \ref{eqn:dynamic_programming} is exactly the same as the CTC dynamic programming \citep{graves-icml-2006}, except we force emit symbols. Equation \ref{eqn:dynamic_programming} also has the added side-effect of having a much lower variance gradient due to the summation term (which can be seen as smoothing out the loss). Figure \ref{fig:dynamic_programming} visualizes the training procedure.

\subsection{Lower bound}
\begin{figure}[t]
    \centering
    \resizebox{0.8\columnwidth}{!}{
    \begin{tikzpicture}[node distance=1cm, scale=1.0, every node/.style={transform shape}]
\tikzstyle{layer} = [rectangle, thick, rounded corners, minimum width=4cm, minimum height=0.5cm, align=center, draw=black]
\tikzstyle{symbol} = [align=center, minimum height=0.6cm]
\tikzstyle{arrow} = [thick,->]

\node (d3) [layer] at (2.5,0) {Softmax};
\node (d2) [layer, below of=d3] {Self-Attention};
\node (d1) [symbol, below of=d2] {$+$};
\node (d0) [layer] at (0, -3.0) {Convolution};
\node (e0) [layer] at (5,-3.0) {Embedding};
\node (audio) [symbol] at (0, -4.0) {$x$: Audio Filter Banks};
\node (hyp) [symbol,align=center] at (5, -4.0) {$\tilde a$: Prior Alignment \\ (contains masked out tokens)};
\node (posterior) [symbol,align=center] at (2.5, 1.25) {New Alignment Posterior \\ $p_\theta(a | x, \tilde a)$};

\draw[arrow] (audio) -- (d0);
\draw[arrow] (d0.north) -- (0, -2.0) -- (d1.west);
\draw[arrow] (d1) -- (d2);
\draw[arrow] (d2) -- (d3);
\draw[arrow] (hyp.north) -- (e0.south);
\draw[arrow] (e0.north) -- (5, -2.0) -- (d1.east);
\draw[arrow] (d3.north) -- (posterior.south);

\end{tikzpicture}
    }
    \caption{Visualization of the Imputer architecture. Imputer conditions on the previous alignment $\tilde a$, a convolutional network processes the audio features $x$, and a Transformer self-attention stack consolidates all available context to generate a new alignment.}
    \label{fig:imputer}
\end{figure}
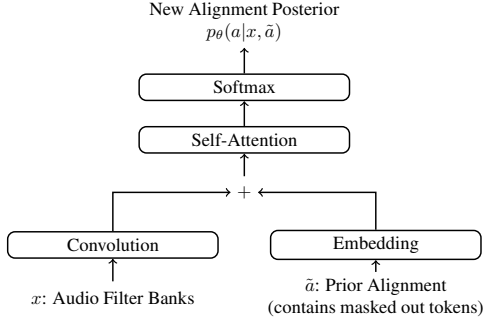

\label{sec:lowerbound}
We can actually show our Imitation Learning algorithm (Equation \ref{eqn:efficient_imitation_learning}) is a lower bound for our Dynamic Programming (Equation \ref{eqn:dynamic_programming}), which in turn is a lower bound for the log probability (up to a constant, which can be dropped during the gradient computation). We start from Equation \ref{eqn:summation}:
{
\begin{align}
\begin{split}
    & \log p_\theta(y | x) \\
    &\quad= \log \sum_{a \in \beta(y)} \sum_{\tilde a \in \gamma(a)} p_\theta(a | \tilde{a}, x) \label{eqn:lowerbound_marginalization}
\end{split} \\
    &\quad= \log \sum_{a \in \beta(y)} \sum_{\tilde a \in \gamma(a)} \sum_{a' \in \beta'(\tilde a, a)} \frac{1}{c(\tilde a, a)} p_\theta(a' | \tilde{a}, x) \label{eqn:lowerbound_doublecount} \\
    &\quad= \log \sum_{a \in \beta(y)} \sum_{\tilde a \in \gamma(a)} \frac{1}{c(\tilde a, a)} \sum_{a' \in \beta'(\tilde a, a)} p_\theta(a' | \tilde{a}, x) \label{eqn:lowerbound_rearrange} \\
    &\quad\geq \mathbb{E}_{a \sim q_{\phi'}} \left[ \mathbb{E}_{\tilde a \sim r} \left[ \log \sum_{a' \in \beta'(\tilde a, a)} p_\theta(a' | \tilde{a}, x) \right] \right] + C \label{eqn:lowerbound_dynamicprogramming}
\end{align}
}%
Equation \ref{eqn:lowerbound_marginalization} writes out the marginalization over all possible alignments $a \in \beta(y)$, and over all possible maskings $\tilde a \in \gamma(a)$. We can rewrite Equation \ref{eqn:lowerbound_marginalization} with an additional summation over $\beta'(\tilde a, a)$, where $\beta'(\tilde a, a)$ captures all compatible alignments with (partially) masked-out alignment $\tilde a$. This will inherently result in repetition of alignments, and the constant $c(\tilde a, a)$ is the number of times each alignment is repeated. We note that the repetition constant $c(\tilde a, a)$ is simply a function of $\tilde a$ and $a$:
\begin{align}
    c(\tilde a, a) = \prod_{s \in \mathrm{masked\_segments}(\tilde a)} \binom{s_n}{s_k}
\end{align}
where $\mathrm{masked\_segments}(\tilde a)$ captures each contiguous masked out segment of $\tilde a$, and $s_n$ is the number of mask tokens in the segment, and $s_k$ is the number of target tokens in the segment. Equation \ref{eqn:lowerbound_rearrange} re-arranges the constant, and 
Equation \ref{eqn:lowerbound_dynamicprogramming} applies importance sampling, Jensen's inequality and collapses the constants into $C = \mathbb{E}_{q_{\phi'}} \left[ H(r) - \mathbb{E}_{r}\left[ \log(c(\tilde a, a)) \right] \right] + H(q_{\phi'})$, since they do not affect the gradient and optimization. Equation \ref{eqn:lowerbound_dynamicprogramming} recovers our dynamic programming algorithm, which is a lower bound up to constant $C$.

It is further obvious that the our Imitation Learning algorithm \ref{eqn:efficient_imitation_learning}) is a lower bound of Equation \ref{eqn:lowerbound_dynamicprogramming}. We replace $a' \in \beta'(\tilde a, a)$ with $a' = a$, which means we are simply dropping terms in the summation; since all the terms dropped are positive, the Imitation Learning algorithm is a (weak) lower bound to our Dynamic Programming algorithm. Note, that both our Imitation Learning and Dynamic Programming algorithm uses the same roll-in policy $\phi' \times \gamma$, which allows us ignore the importance sampling correction terms.

\section{Model}

In this section, we describe the neural network architecture of our Imputer implementation. Figure \ref{fig:imputer} visualizes the Imputer neural network. The Imputer processes the $x$ audio features (e.g., filter banks) with a stack of convolutional layers. These features are then added to alignment embeddings (from a previous alignment $\tilde a$), which are then passed through a stack of Transformer self-attention layers \citep{vaswani-nips-2017}. Finally, a softmax layer models the new alignment $a$. We describe the exact hyperparameters used in Section \ref{sec:experiments}.

\section{Experiments}
\label{sec:experiments}

\begin{figure*}[t]
    \centering
    \tiny
    \setlength{\tabcolsep}{4pt}
    \begin{tabular}{c|cccccccc|cccccccc|cccccccc|cccccccc}
        \bfseries $b$ & \multicolumn{32}{c}{\bfseries Alignment} \\
        \midrule
        0 & $\mask$ & $\mask$ & $\mask$ & $\mask$ & $\mask$ & $\mask$ & $\mask$ & $\mask$ & 
        $\mask$ & $\mask$ & $\mask$ & $\mask$ & $\mask$ & $\mask$ & $\mask$ & $\mask$ & $\mask$ & $\mask$ & $\mask$ & $\mask$ & $\mask$ & $\mask$ & $\mask$ & $\mask$ & $\mask$ & $\mask$ & $\mask$ & $\mask$ & $\mask$ & $\mask$ & $\mask$ & $\mask$ \\

        1 & $\mask$ & $\mask$ & $\mask$ & [HAVE & $\mask$ & $\mask$ & $\mask$ & $\mask$ & 
        $\mask$ & [L & $\mask$ & $\mask$ & $\mask$ & $\mask$ & $\mask$ & $\mask$ &
        $\mask$ & $\mask$ & $\mask$ & Y & $\mask$ & $\mask$ & $\mask$ & $\mask$ &
        $\mask$ & $\mask$ & L & $\mask$ & $\mask$ & $\mask$ & $\mask$ & $\mask$ \\

        2 & $\mask$ & $\mask$ & $\mask$ & [HAVE & $\mask$ & $\mask$ & [TO & $\mask$ & 
        $\mask$ & [L & $\mask$ & $\mask$ & IVE & $\mask$ & $\mask$ & $\mask$ &
        $\mask$ & $\mask$ & $\mask$ & Y & $\mask$ & $\mask$ & [SE & $\mask$ &
        $\mask$ & $\mask$ & L & $\mask$ & F & $\mask$ & $\mask$ & $\mask$ \\

        3 & $\mask$ & $\mask$ & $\_$ & [HAVE & $\mask$ & $\mask$ & [TO & $\mask$ & 
        $\_$ & [L & $\mask$ & $\mask$ & IVE & $\mask$ & $\mask$ & $\mask$ &
        $\mask$ & [M & $\mask$ & Y & $\mask$ & $\mask$ & [SE & $\mask$ &
        $\_$ & $\mask$ & L & $\mask$ & F & $\mask$ & $\mask$ & $\mask$ \\

        4 & $\mask$ & $\mask$ & $\_$ & [HAVE & $\mask$ & $\mask$ & [TO & $\_$ & 
        $\_$ & [L & $\mask$ & $\mask$ & IVE & $\_$ & $\mask$ & $\mask$ &
        $\mask$ & [M & $\mask$ & Y & $\mask$ & $\mask$ & [SE & $\_$ &
        $\_$ & $\mask$ & L & $\mask$ & F & $\mask$ & $\mask$ & $\_$ \\

        5 & $\_$ & $\mask$ & $\_$ & [HAVE & $\mask$ & $\mask$ & [TO & $\_$ & 
        $\_$ & [L & $\mask$ & $\_$ & IVE & $\_$ & $\mask$ & $\mask$ &
        $\mask$ & [M & $\_$ & Y & $\mask$ & $\mask$ & [SE & $\_$ &
        $\_$ & $\_$ & L & $\mask$ & F & $\mask$ & $\mask$ & $\_$ \\

        6 & $\_$ & $\mask$ & $\_$ & [HAVE & $\mask$ & $\_$ & [TO & $\_$ & 
        $\_$ & [L & $\mask$ & $\_$ & IVE & $\_$ & [WITH & $\mask$ &
        $\mask$ & [M & $\_$ & Y & $\mask$ & $\_$ & [SE & $\_$ &
        $\_$ & $\_$ & L & $\mask$ & F & $\mask$ & $\_$ & $\_$ \\

        7 & $\_$ & $\_$ & $\_$ & [HAVE & $\mask$ & $\_$ & [TO & $\_$ & 
        $\_$ & [L & $\_$ & $\_$ & IVE & $\_$ & [WITH & $\mask$ &
        $\_$ & [M & $\_$ & Y & $\mask$ & $\_$ & [SE & $\_$ &
        $\_$ & $\_$ & L & $\_$ & F & $\mask$ & $\_$ & $\_$ \\

        8 & $\_$ & $\_$ & $\_$ & [HAVE & $\_$ & $\_$ & [TO & $\_$ & 
        $\_$ & [L & $\_$ & $\_$ & IVE & $\_$ & [WITH & $\_$ &
        $\_$ & [M & $\_$ & Y & $\_$ & $\_$ & [SE & $\_$ &
        $\_$ & $\_$ & L & $\_$ & F & $\_$ & $\_$ & $\_$ \\
    \end{tabular}
    \caption{Example Imputer inference from the LibriSpeech dev set with block size $B=8$ for the target sequence ``[HAVE [TO [LIVE [WITH [MYSELF'' with generated alignment ``\_ \_ \_ [HAVE \_ \_ [TO \_ \_ [L \_ \_ IVE \_ [WITH \_ \_ [M \_ Y \_ \_ [SE \_ \_ \_ L \_ F \_ \_ \_ ''. Imputer takes exactly block size number of generation steps ($B=8$) to generate the entire sequence, independent of the output sequence length. At each $b$ iteration, one token is filled in within each block and with parallel generation across blocks.}
    \label{fig:imputation}
\end{figure*}

\begin{table}[t]
\centering
\caption{Wall Street Journal Character Error Rate (CER) and Word Error Rate (WER).}
\label{tab:wsj}
\scalebox{0.8}{
\begin{tabular}{lccc}
 \toprule
 \textbf{Model} & \textbf{CER} & \textbf{WER} & \bfseries Iterations \\ 
 \midrule
 seq2seq \\
 \quad \citet{bahdanau-icassp-2016} & 6.4 & 18.6 & n \\
 \quad \citet{bahdanau-iclr-2016} & 5.9 & 18.0 & n \\
 \quad \citet{chorowski-interspeech-2017} & - & 10.6 & n \\
 \quad \citet{zhang-icassp-2017} & - & 10.5 & n \\
 \quad \citet{chan-iclr-2017} & - & 9.6 & n \\
 \quad \citet{kim-icassp-2017} & 7.4 & - & n \\
 \quad \citet{serdyuk-iclr-2018} & 6.2 & - & n \\
 \quad \citet{tjandra-icassp-2018} & 6.1 & - & n \\
 \quad \citet{sabour-iclr-2019} & 3.1 & 9.3 & n \\
 \midrule
 CTC \\

 \quad \citet{graves-icml-2014} & 8.4 & 27.3 & 1 \\
 \quad \citet{liu-icml-2017} & - & 16.7 & 1 \\
 \quad CTC (Our Work) & 5.6 & 15.2 & 1 \\
 \midrule
 Imputer (IM) & 6.2 & 16.5 & 8 \\
 Imputer (DP) & 4.9 & 12.7 & 8 \\
 \bottomrule
\end{tabular}
}
\end{table}

\begin{table}[t]
\centering
\caption{LibriSpeech test-clean and test-other Word Error Rate (WER).}
\label{table:librispeech}
\scalebox{0.8}{
\begin{tabular}{lccc}
\toprule
\bfseries Method & \bfseries clean & \bfseries other & \bfseries Iterations \\
\midrule
seq2seq \\
\quad \citet{zeyer-interspeech-2018} & 4.9 & 15.4 & n \\
\quad \citet{zeyer-nips-2018} & 4.7 & 15.2 & n \\
\quad \citet{irie-arxiv-2019} & 4.7 & 13.4 & n \\
\quad \citet{sabour-iclr-2019} & 4.5 & 13.3 & n \\
\quad \citet{luscher-interspeech-2019} & 4.4 & 13.5 & n \\
\quad \citet{park-interspeech-2019} & 4.1 & 12.5 & n \\
\midrule
ASG/CTC \\
\quad \citet{collobert-arxiv-2016} & 7.2 & - & 1 \\
\quad \citet{liptchinsky-arxiv-2017} & 6.7 & - & 1 \\
\quad CTC (Our Work) & 4.6 & 13.0 & 1 \\
\midrule
Imputer (IM) & 5.5 & 14.6 & 8 \\
Imputer (DP) & 4.0 & 11.1 & 8 \\
\bottomrule
\end{tabular}
}
\end{table}

\begin{figure}[t]
    \centering
    \resizebox{.66\linewidth}{!}{{\pgfplotsset{compat=1.3}
\begin{tikzpicture}
\begin{axis}[
    title={Training/Inference Block Size and Decoding Strategy vs WER},
    xlabel={Block Size},
    ylabel={Word Error Rate},
    ymin=10.5, ymax=14.5,
    xmin=0, xmax=6,
    xticklabels={2,4,8,16,32}, xtick={1,2,3,4,5},
    ytick={11,12,13,14},,
    ymajorgrids=true,
    grid style=dashed,
    legend pos=north east,
    legend cell align=left
]
\addplot[
    color=blue,
    mark=*
    ]
    coordinates {
    (1,14.2)(2,11.3)(3,11.3)(4,11.8)(5,12.7)
    };
\addplot[
    color=red,
    mark=square*,
    very thick,
    dotted,
    mark options={solid}
    ]
    coordinates {
    (1,14.2)(2,11.7)(3,11.4)(4,11.7)(5,12.1)
    };
    \legend{Alternate Sub-block,
    Right Most Last}
\end{axis}
\end{tikzpicture}}}
    \caption{LibriSpeech dev-other WER with different block size used for training/inference and decoding strategy.}
    \label{fig:wer_vs_bs_train}
\end{figure}

\begin{figure}[t]
    \centering
    \resizebox{.66\linewidth}{!}{{\pgfplotsset{compat=1.3}
\begin{tikzpicture}
\begin{axis}[
    title={Inference Block Size and Decoding Strategy vs WER},
    xlabel={Block Size},
    ylabel={Word Error Rate},
    ymin=10.5, ymax=15.5,
    xmin=0, xmax=9,
    xticklabels={2,4,8,16,32,64,128,256}, xtick={1,2,3,4,5,6,7,8},
    ytick={11,12,13,14,15},
    legend pos=north west,
    ymajorgrids=true,
    grid style=dashed,
    legend cell align=left
]
\addplot[
    color=blue,
    mark=*
    ]
    coordinates {
    (1,13.6)(2,11.4)(3,11.3)(4,11.13)(5,11.4)(6,12)(7,13.1)(8,15.0)
    };
\addplot[
    color=red,
    mark=square*,
    very thick,
    dotted,
    mark options={solid}
    ]
    coordinates {
    (1,13.6)(2,11.8)(3,11.4)(4,11.14)(5,11.1)(6,11.05)(7,11.03)(8,11)
    };
    \legend{Alternate Sub-block,
    Right Most Last}
\end{axis}
\end{tikzpicture}}}
    \caption{LibriSpeech dev-other WER with different block size used for inference and decoding strategy trained with block size 8.}
    \label{fig:wer_vs_bs_decode}
\end{figure}

We experiment with two competitive speech tasks, the 82 hours Wall Street Journal (WSJ) \citep{paul-snl-1992} dataset and the 960 hours LibriSpeech \citep{panayotov-icassp-2015} dataset. We use Kaldi \citep{povey-asru-2011} to generate 80-dimensional filter banks, with delta and delta-delta. We use SentencePiece \citep{kudo-emnlp-2018} to generate a 400 BPE subword vocabulary. We compare our Imputer models to other end-to-end speech recognition networks (i.e., without relying on an external language model). We also opt to compare results without the use of data augmentation, since different authors use different data augmentation methods (e.g., speed perturbation \citep{li-interspeech-2019}, tempo/volume perturbation \citep{tuske-interspeech-2019}, SpecAugment \citep{park-interspeech-2019}, etc...) which makes fair comparison difficult.

Our neural network uses 2 layers of convolution each with $11 \times 3$ filter size and stride of $2 \times 1$. For our WSJ experiments, we use 8 Transformer self-attention layers with 4 attention heads, 512 hidden size, 2048 filter size, dropout rate 0.2 and train for 300k steps. For our LibriSpeech experiments, we use 16 Transformer self-attention layers with 4 attention heads, 512 hidden size, 4096 filter size, dropout rate 0.3 and train for 1M steps. We perform no model selection and simply report the CER/WER at the end of training.

We built strong CTC baselines with the architecture described above, which outperform prior non-autoregressive work. We also use our CTC models as our expert alignment model $\phi$ for our alignment roll-in. We add a small amount of noise by shuffling the alignment by up to 1 frame to the left-or-right, however this was not critical and only resulted in very minor improvements (\textless 0.2) in WER. Our Imputer results are reported with the Block masking policy.

Table \ref{tab:wsj} and Table \ref{table:librispeech} summarizes our WSJ and LibriSpeech experiments respectively, along with prior work. We find our Imputer model to outperform prior non-autoregressive work. For WSJ, our Imputer model achieves 12.7 WER, while our non-autoregressive CTC baseline achieves 15.2 WER. We find our Imputer model competitive to prior work of autoregresive seq2seq models which achieve 9.3 WER \citep{sabour-iclr-2019}. For LibriSpeech, our CTC baseline achieves 13.0 WER, a strong autoregressive seq2seq baseline achieves 12.5 WER \citep{park-interspeech-2019}, and our Imputer model achieves 11.1 WER outperforming both works. Figure \ref{fig:imputation} gives an example decoding path from LibriSpeech.

We find our Dynamic Programming (DP) algorithm to outperform our Imitation Learning (IL) algorithm in both WSJ and LibriSpeech. Our IL algorithm achieves 14.6 WER while our DP algorithm achieves 11.1 WER for LibriSpeech. This is expected since our DP algorithm is a tighter lower bound for the likelihood. However, we also found our IL algorithm to perform worse than our CTC baseline. One hypothesis is that the variance of the IL gradient may be much higher, and the usage of dynamic programming may help in both generalization and optimization.

\subsection{Analysis}
In this section we will analyze different decoding strategies and training/inference block sizes.

\textbf{Decoding Strategy.}
During inference, we want to avoid making local conditional independence assumptions. We want to avoid neighbouring tokens to be generated in parallel. Our inference strategy to $\argmax$ only 1 token per block per generation iteration mostly accomplishes this. However, there is the edge case between the block boundaries which can result in two neighbouring tokens generated in parallel at the same time. We present two decoding strategies that will alleviate the model from generating parallel neighbouring tokens at this boundary condition:
\begin{enumerate}[nosep]
    \item \textbf{Alternate Sub-block}. Each block is divided into two sub-blocks, left sub-block and right sub-block. On even iterations, only the left sub-block is allowed to be imputed, and during odd iterations, only the right sub-block is allowed to be imputed.
    \item \textbf{Right Most Last}. The right-most token in each block is only permitted to be imputed at the last iteration.
\end{enumerate}

\textbf{Block Size.}
The block size is an important hyperparameter for the Imputer model, since it trades-off between inference speed and model contextualization.
A small block size will result in a smaller number of generation iterations, while a large block size will result in fewer conditional independence assumptions.
We analyze the effect of block sizes in two ways:
\begin{enumerate}[nosep]
    \item We train our model to various block sizes $B \in \{2, 4, 8, 16, 32\}$, and use the same block size for inference.
    \item We train our model to a fixed block size $B=8$, and use different block sizes for inference.
\end{enumerate}
We report both the decoding strategy and block size experiments for the LibriSpeech dev-other split in Figure \ref{fig:wer_vs_bs_train} and Figure \ref{fig:wer_vs_bs_decode}.
Figure \ref{fig:wer_vs_bs_train} compares the performance of the models trained with different block sizes (with inference using the same block size they were trained on), along with the two different decoding strategies. In case of both decoding strategies, block size $B=8$ gives the lowest WER. As we decrease the block size further, WER increases because of more conditionally independent (parallel) generations. Interestingly, we find that models trained and decoded with larger block sizes also yield worse WER, even though they allow more conditional dependencies. We hypothesize this could be an optimization issue, as we found it more difficult to train larger block size models (e.g., the $\beta \times \gamma$ space is larger for larger block sizes).

Figure \ref{fig:wer_vs_bs_decode} compares the performance where the model is trained with a block size of $B=8$, but decoded with various block sizes. When using Right Most Last decoding strategy, we find a monotonic decrease in WER with increase in block size resonating with the hypothesis that more conditional dependency should result in better WERs at the cost of inference speed. However, when using Alternate Sub-block decoding strategy, we observe a local minima at around $B=16$. We hypothesize that this behaviour occurs due to the over-constraining nature of Alternate Sub-block decoding strategy, as it forces the model to predict tokens for slots in specific sub-blocks at each iteration.

\subsection{Masking Policy}
As discussed in Section \ref{sec:rollin_policy}, we can use different masking policy $\gamma$. Table \ref{tab:masking} reports the results over 3 different masking policies: Bernoulli, Uniform and Block. We find Imputer to be robust to the choice of masking policy.

\begin{table}[t]
\centering
\caption{LibriSpeech dev-other WER for various masking policies.}
\label{tab:masking}
\scalebox{0.8}{
\begin{tabular}{lc}
\toprule
\bfseries Masking Policy & \bfseries WER \\
\midrule
Bernoulli & 11.2 \\
Uniform & 11.4 \\
Block & 11.4 \\
\bottomrule
\end{tabular}
}
\end{table}

\subsection{Training with Model Samples}
In Section \ref{sec:rollin_policy}, we discussed sampling from a (stale) copy of our model for the alignments in the roll-in policy. We trained models where we started off with roll-in alignments from a pre-trained CTC model (with noise), then decoded our Imputer model every 50k steps for our new alignments. However, we did not find improvements in performance, suggesting that using the CTC alignment is a sufficient for the roll-in alignments to train Imputer.

\subsection{Top-$k$ Decoding}
We can ignore the block structure during inference, and decode via greedy top-k. This will once again result in a model requiring only constant $B$ generation steps if we choose $k = \sfrac{|x|}{B}$. Once again, care is taken such that we do not impute neighbouring tokens in parallel. We find our models to perform slightly worse under this condition, suggesting the block inference strategy to reasonable, especially since speech is relatively local. We decode our Bernoulli model on LibriSpeech dev-other and it achieves 11.6 WER, compared to 11.2 WER using our Block inference strategy.

\subsection{Simulated Annealing Decoding}
One alternative view of Imputer is to view it as a transition operator in an energy-based system. We can use a simulated annealing like procedure for decoding. We tried various strategies where we deleted token commitments via a random sampling policy or based on the token-level likelihood (similar to Mask-Predict \cite{ghazvininejad-emnlp-2019}). We found improvements for unconverged models, however we did not find improvements on our converged models.
\section{Related Work}
Non-Autoregressive Transformer (NAT) \citep{gu-iclr-2018} and Mask-Predict \citep{ghazvininejad-emnlp-2019} are closely related to our Imputer model. Mask-Predict was also recently adapted for end-to-end speech recognition \citep{chen-arxiv-2019}. There are several key differences, 1) these models require predicting the target sequence length first, and 2) these models follow an encoder-decoder formulation which requires a cross-attention mechanism that may not be well suited for long monotonic sequence problems like speech recognition. Imputer does not rely on length prediction first but relies on a fixed size canvas, and the local prediction structure of the Imputer architecture is perhaps more suited for monotonic problems like speech recognition.

The Insertion Transformer \citep{stern-icml-2019} and other insertion-based models \citep{gu-arxiv-2019,welleck-icml-2019,chan-arxiv-2019,li-wngt-2019,ruis-wngt-2019} generates sequences through a series of insertions. The Insertion Transformer can dynamically grow its canvas size, whereas the Imputer has a fixed canvas size. Insertion Transformer has been shown to generate sequences in logarithmic $\log n$ iterations \citep{stern-icml-2019}, where as the Imputer requires constant generation steps. Additionally, the Insertion Transformer with its vastly non-monotonic generation order \citep{chan-arxiv-2019b} lacks the monoticity that the Imputer dynamic programming endows, consequently we found the Insertion Transformer difficult to apply to speech recognition applications.

Our neural network architecture was also inspired by past work in convolutions \citep{sainath-icassp-2013} and Transformers \citep{dong-icassp-2018} for speech recognition. \citet{mohamed-arxiv-2019} also combined convolutions with self-attention, where they also use convolutions to model local acoustic features, and uses self-attention to capture global signals from both the acoustics and the previous alignment, which licenses powerful bidirectional language modelling.
Finally, we also mention the concurrent work of \citet{synnaeve-arxiv-2019}, where they also apply a convolutional Transformer architecture similiar to ours, showing strong empirical results on both CTC and seq2seq models. However, their work involved substantially larger models with data augmentation and focus on semi-supervised learning.
\section{Conclusion}

In this paper, we presented the Imputer, a neural sequence model which generates sequences iteratively via imputations. Imputer iteratively conditions on a previous partial alignment, and generates a new alignment. Imputer requires only a fixed constant number of generative iterations independent of the number of input or output tokens. Imputer can be trained to approximately marginalize over all possible alignments and generation order. We presented a tractable dynamic programming training algorithm. Our dynamic program marginalizes over all possible completions given any previous alignment, and we show it to be a lower bound for the likelihood. We apply Imputer to end-to-end speech recognition tasks, and find Imputer outperforming prior non-autoregressive approaches and achieving comparable performance to autoregressive models. On LibriSpeech test-other, Imputer achieves 11.1 WER, outperforming CTC at 13.0 WER and seq2seq at 12.5 WER.

\section*{Acknowledgements}
We are grateful to Simon Kornblith, Sara Sabour, Saurabh Saxena, Yu Zhang and the Google Brain team for technical assistance and discussions.

\bibliography{paper}
\bibliographystyle{icml2020}

\end{document}